\documentclass[sigconf,nonacm]{acmart} %
\AtBeginDocument{%
  }

\copyrightyear{2025}
\acmYear{2025}
\setcopyright{rightsretained}
\acmConference[CHI EA '25]{Extended Abstracts of the CHI Conference on Human Factors in Computing Systems}{April 26-May 1, 2025}{Yokohama, Japan}
\acmBooktitle{Extended Abstracts of the CHI Conference on Human Factors in Computing Systems (CHI EA '25), April 26-May 1, 2025, Yokohama, Japan}\acmDOI{10.1145/3706599.3719699}
\acmISBN{979-8-4007-1395-8/2025/04}

\usepackage{acmart-taps}
\usepackage{enumitem}

\makeatletter
\newcommand\newtag[2]{#1\def\@currentlabel{#1}\label{#2}}
\makeatother

\begin{document}

\title[What Makes a Model Breathe? Understanding Reinforcement Learning Reward Function Design]{What Makes a Model Breathe? Understanding Reinforcement Learning Reward Function Design in Biomechanical User Simulation}

\author{Hannah Selder}
\orcid{0009-0008-7049-1630} 
\affiliation{%
  \institution{Center for Scalable Data Analytics and Artificial Intelligence (ScaDS.AI), Leipzig University}
  \city{Leipzig}
  \country{Germany}
}
\email{hannah.selder@uni-leipzig.de}

\author{Florian Fischer}
\orcid{0000-0001-7530-6838} 
\affiliation{%
  \institution{University of Cambridge}
  \city{Cambridge}
  \country{United Kingdom}}
\email{fjf33@cam.ac.uk}

\author{Per Ola Kristensson}
\orcid{0000-0002-7139-871X} 
\affiliation{%
  \institution{University of Cambridge}
  \city{Cambridge}
  \country{United Kingdom}}
\email{pok21@cam.ac.uk}

\author{Arthur Fleig}
\orcid{0000-0003-4987-7308} 
\affiliation{%
  \institution{Center for Scalable Data Analytics and Artificial Intelligence (ScaDS.AI), Leipzig University}
  \city{Leipzig}
  \country{Germany}
}
\email{arthur.fleig@uni-leipzig.de}

\renewcommand{\shortauthors}{Selder et al.}

\begin{abstract} %
    Biomechanical models allow for diverse simulations of user movements in interaction.
    Their performance depends critically on the careful design of reward functions, yet the interplay between reward components and emergent behaviours remains poorly understood. 
    We investigate what makes a model “breathe” by systematically analysing the impact of rewarding effort minimisation, task completion, and target proximity on movement trajectories. 
    Using a choice reaction task as a test-bed, we find that a combination of completion bonus and proximity incentives is essential for task success. Effort terms are optional, but can help avoid irregularities if scaled appropriately.
    Our work offers practical insights for HCI designers to create realistic simulations without needing deep reinforcement learning expertise, advancing the use of simulations as a powerful tool for interaction design and evaluation in HCI.
\end{abstract}



\maketitle

\section{Introduction}
Biomechanical simulations offer great potential for modelling and understanding human body movements.
They can be used to predict muscle fatigue~\cite{cheema_predicting_2020}, anticipate potential injury risks~\cite{wan2016biomechanical}, and guide the design of rehabilitation programs~\cite{sacco_2023}.
In the context of human-computer interaction (HCI), biomechanical models have been successful in predicting how users type on a virtual keyboard~\cite{hetzel2021complex}, point at and track moving mid-air targets~\cite{fischer2021reinforcement, ikkala_breathing_2022, klar_simulating_2023}, or play exergames in virtual reality (VR)~\cite{fischer_sim2vr_2024}.
Recently, visual perception models have been integrated with musculoskeletal dynamics~\cite{ikkala_breathing_2022}, strengthening the value of biomechanical simulations as a basis for comprehensive user models.

However, the quality of biomechanical user simulations essentially relies on their ability to reliably capture task-relevant behaviours and motion patterns.
In reinforcement learning (RL), the most promising and commonly used technique to forward-simulate musculoskeletal models~\cite{ikkala_breathing_2022, caggiano2022myosuite, song2021deep}, a key ingredient is an appropriate \textit{reward function}.
RL (and other optimal control methods such as Model Predictive Control~\cite{klar_simulating_2023} or the Linear Quadratic Gaussian Control~\cite{fischer_optimal_2022}) is based on the assumption of rational behaviour:
an \textit{agent}, that is, a real or simulated entity capable of making decisions within a defined context, is assumed to observe their environment and %
ensure their actions align with their overarching goals. %
The reward function summarises the agent's goals, and therefore may change as the task or context changes. For example, if the agent is to grasp a cup, or type a certain word on a keyboard, this should be reflected in the reward function.

While recent works have demonstrated the potential of RL-trained agents to simulate plausible human movement in various HCI contexts, e.g.,~\cite{fischer2021reinforcement,ikkala_breathing_2022}, they rarely communicate the amount of work spent tuning the reward function.
Depending on the task's complexity and specificity, identifying an appropriate reward function can be very challenging.
Main challenges include the lack of general guidelines for novices, missing insights into the comparative performance of relevant reward components, and the difficulty of ``trial-and-error'' approaches due to extensive RL training times when using state-of-the-art biomechanical models, which are typically in the order of 12--72 hours on modern workstations for testing a single reward function. 
This considerably limits the applicability of current biomechanical simulations to HCI and prevents further engagement with RL-based simulation frameworks.

This late-breaking work constitutes a first key step toward addressing current issues related to reward function design in biomechanical user simulations.
We specifically consider the choice reaction task from~\cite{ikkala_breathing_2022}, which is well-established in HCI.
In this task, users are given several buttons and a monitor in front of them and need to click the button of the colour currently displayed on the monitor as fast as possible. This task requires non-trivial muscle coordination and skills essential to visuomotor interaction: colour vision, object recognition and aimed arm movements. 
For this choice reaction task, we provide a comprehensive and thorough evaluation of relevant reward terms through 60 trained policies. %
In particular, we address the following \textbf{research questions}: 
\aptLtoX{\begin{enumerate}%
		\item[\textbf{RQ.1}]\label{item:rq:plausible-movement} \textbf{(Plausibility)} What combinations and relative weightings of reward function components (e.g., proximity, effort, and task completion bonuses) produce plausible human movement trajectories %
    in a choice reaction task?
		\item[\textbf{RQ.2}]\label{item:rq:sensititivies} \textbf{(Sensitivity)} What are the sensitivities of interaction outcomes %
    to variations in individual reward function components in a choice reaction task?
	\end{enumerate}}{
\begin{enumerate}[label=\textbf{RQ.\arabic*}]
    \item\label{item:rq:plausible-movement} \textbf{(Plausibility)} What combinations and relative weightings of reward function components (e.g., proximity, effort, and task completion bonuses) produce plausible human movement trajectories %
    in a choice reaction task?
    \item\label{item:rq:sensititivies} \textbf{(Sensitivity)} What are the sensitivities of interaction outcomes %
    to variations in individual reward function components in a choice reaction task?
\end{enumerate}}
With \ref{item:rq:plausible-movement} we aim to help researchers and practitioners who seek optimized reward functions. With \ref{item:rq:sensititivies} we aim to gain a better understanding of the difficulties inherent to reward function design and to derive guidelines for the development of composite reward functions.

Specifically, we \textbf{contribute to computational modelling of movement-based interaction} by providing
\begin{itemize}
    \item a first systematic exploration of standard reward function components for a movement-based HCI task, and 
    \item guidelines and first principles for reward function design in RL-based biomechanical user simulations.
\end{itemize}

\section{Related Work}

Biomechanical simulations are a beneficial tool for developing and validating HCI technologies~\cite{murray-smith_what_2022, bachynskyi2015performance, fischer_sim2vr_2024}. 
They utilize dynamic models of the human body to predict movement during interaction. 
While early models were limited to calculating mechanical loads in static postures~\cite{winter_biomechanics_1984, flash_coordination_1985, Ayoub_Biomechanical_1974}, advancements have led to physiologically increasingly accurate musculoskeletal models~\cite{madeleine_biomechanics_2011, damsgaard_analysis_2006, delp_opensim_2007}.
These models are integrated into user simulations to generate realistic movements~\cite{roupa2022modeling, millard2019reduced}.
For example, 
\citeauthor{hwang_ergopulse_2024} apply simulations with electrical muscle stimulation to create kinesthetic force feedback for virtual reality~\cite{hwang_ergopulse_2024}.
Biomechanical models have also been used to simulate human movements in interaction tasks~\cite{fischer_optimal_2022, klar_simulating_2023, ikkala_breathing_2022,fischer_sim2vr_2024} and to analyse the cognitive aspects of interaction tasks~\cite{chen_cognitive_2017, chen_emergence_2015}.

Deep RL has emerged as the go-to method for simulating movement-based interaction. For example, \citeauthor{fischer2021reinforcement} use it to learn controlling the muscles of a state-of-the-art shoulder model for a mid-air pointing task.
\citeauthor{ikkala_breathing_2022} present \textit{User-in-the-Box}, an RL-based simulation framework to generate task-specific movement trajectories based on the user's visual and proprioceptive perception of the interaction environment.
Other optimal feedback control methods have been investigated for simulating movement-based interaction \cite{fischer_optimal_2022, klar_simulating_2023, martin_intermittent_2021}. 
However, these have shown applicable for relatively low-dimensional %
control problems only, imposing severe restrictions on the complexity of the biomechanical models and tasks considered.

In RL-based biomechanical simulations, the design of the reward function is identified as a key factor in the effectiveness of the learning process \cite{kwiatkowski_reward_2023}.
Consequently, formulations of effective reward functions for specific simulated tasks have been examined \cite{kwiatkowski_reward_2023, he_exploring_2024, nowakowski_human_2021,ikkala_breathing_2022}. %
However, most of the proposed reward functions have been handcrafted for a specific task~\cite{caggiano2023myodex, caggiano2022myosuite, ikkala_breathing_2022}, limiting their generalizability across tasks and contexts.
In addition, reward functions usually involve a trade-off between two or more opposing objectives, e.g., between accuracy and stability~\cite{liu2007evidence} or speed and accuracy~\cite{nagengast2011risk}. In particular, composite reward functions typically include at least one ``effort'' term that penalizes large controls, which restricts the use of rapid and abrupt arm movements and ensures that available resources are used efficiently. %
Several effort cost models have been proposed and investigated from a motor control perspective~\cite{berret_evidence_2011, wang_advances_2011, wada_quantitative_2001, cheema_predicting_2020}.
These models address the redundancy of movement problem, which refers to the fact that humans can perform tasks with an infinite number of different admissible joint trajectories~\cite{berret_evidence_2011}. This is because penalizing different behaviours, such as rapid and jerky arm movements, leads to different movement patterns~\cite{berret_evidence_2011, guigon_computational_2007}.
Furthermore, the role of an effort term in motor adaptation is explored in the empirical studies in~\cite{xu_inducing_2021, proietti_modifying_2017}.

While different effort cost models have been proposed and investigated from a motor control perspective~\cite{berret_evidence_2011, wang_advances_2011, wada_quantitative_2001, cheema_predicting_2020}, there exist no guidelines on how to design and balance these reward components in practical HCI tasks, especially in combination with complex musculoskeletal systems.

We therefore anticipate a strong need to explore the design of reward functions for realistic use cases of biomechanical models. In this work, we make a decisive step towards this goal by starting with a simple choice reaction task and analysing the individual and combined effects of different reward function components on RL-based learning of interactive body movement.

\section{Methodology}
In this work, we analyse the effect of different reward components on the predicted user strategies in an RL-based biomechanical simulation approach.
We focus on the choice reaction task implemented in the User-in-the-Box (UitB) framework\footnote{\url{https://github.com/User-in-the-Box/user-in-the-box}}.
The agent is provided with four different coloured buttons and a stimulus (one of the four colours) shown on a display in front of them. The task is to press the button of the displayed colour as fast as possible, within a maximum period of four seconds per trial. As soon as the correct button is pressed with a suitable force, the displayed colour switches and the next trial starts.

For each considered reward function, we train an RL policy within the UitB choice reaction environment following the procedure described in~\citet{ikkala_breathing_2022}. In particular, we use the default \textit{MoBL Arms Model}~\cite{saul2015benchmarking} with 5 DoFs (three independent shoulder joints, elbow, wrist) and 26 muscles enabled, and provide visual, proprioceptive, and tactile information as input to the agent.
Each episode starts with the arm hanging down (see Figure~\ref{fig:effort_models} (left)).
Each policy trained for a given reward function can be used to simulate and predict user behaviour; we therefore denote a trained agent as \textit{simulated user} in the following.

When designing the reward function, we focus on three components:
\begin{itemize}
    \item The \textit{completion bonus} component rewards task completion, e.g., similar to scores in games. 
    While it can be simply a constant for tapping the right button, we integrate the many possibilities in the function $f_{\text{bonus}}(\cdot)$, where $(\cdot)$ is a placeholder for all relevant function arguments.
    \item The \textit{distance} component rewards the agent more the closer they get to the target. 
    In the choice-reaction task, this means moving towards the right-coloured button. 
    To incorporate the many mathematical formulations, we introduce the function $f_{\text{distance}}(\cdot)$.
    \item The \textit{effort} component is very versatile: 
    Designers can choose, e.g., to penalize jerky movements, or reward movements that require lower energy. 
    We encompass the possibilities in the function $f_{\text{effort}}(\cdot)$. 
\end{itemize}
To evaluate the intricacies of how the individual components work independently and in conjunction, for each component we introduce respective weights $w_{\text{distance}}, w_{\text{effort}}, w_{\text{bonus}} \geq 0$. 
In total, the most generic reward function amounts to 
\begin{equation}\label{eq:reward-fct-composite}
    r_t = w_{\text{bonus}} \cdot f_{\text{bonus}}(\cdot) - w_{\text{distance}} \cdot f_{\text{distance}}(\cdot) - w_{\text{effort}} \cdot f_{\text{effort}}(\cdot).
\end{equation}
If we set $w_{\text{effort}}=0$ or $w_{\text{distance}}=0$, we speak of \textit{zero effort} or \textit{zero distance}, respectively.

\subsection{Reward Components}

Each simulated user was trained for 35M steps, %
as we observed that further training beyond this point did not yield additional learning or improvements (the UitB framework suggests a default of 50M training steps).

For the \textit{completion bonus}, we follow~\cite{ikkala_breathing_2022} and consider different constant values~$b \geq 0$, i.e., 
\begin{equation}\label{eq:bonus}
    f_\text{bonus}(\cdot) = \begin{cases}
        b, & \text{if the correct button is pressed, } \\
        0& \text{else}.
    \end{cases}\tag{Bonus}
\end{equation}

We investigate three different \textit{distance} reward functions, each based on the distance between the index finger and the surface of the currently desired button, $dist$, as measured by a MuJoCo distance sensor~\cite{todorov_mujoco_12}:

\begin{enumerate}
    \item The (absolute) value of the MuJoCo distance sensor: %
    \begin{equation} \label{eq:absolute_dist}
        f_{\text{distance}}(dist) = |dist|  \tag{$D_\text{absolute}$}
    \end{equation}
    \item The squared distance, which has been successfully used in RL tasks~\cite{nocedal_numerical__2006, kim_approach_2023}:%
    \begin{equation} \label{eq:squared_dist}
        f_{\text{distance}}(dist) = dist^2 \tag{$D_\text{squared}$}
    \end{equation}
    \item An exponential transformation of the distance, as used in \cite{ikkala_breathing_2022}: 
    \begin{equation} \label{eq:exp_dist}
        f_{\text{distance}}(dist) = \frac{1 - e^{-10\cdot dist}}{10}  \tag{$D_\text{exponential}$}
    \end{equation}
\end{enumerate}

We also compare different \textit{effort} models.
The first one, denoted as \textit{EJK} in the following, was first presented in \cite{charaja_generating_2024} to simulate realistic arm movements and consists of three components. 
This is motivated by the observation that combining multiple effort terms can improve the plausibility of generated movements~\cite{berret_evidence_2011, wochner_optimality_2020}. Its components penalize the mean value of the muscle stimulation commands ($r_\text{energy}$), the jerk, i.e., the change in joint acceleration ($r_\text{jerk}$), and the total work done by the shoulder and elbow ($r_\text{work}$) in terms of angular velocities and torques. 
These components are normalized and weighted by coefficients $c_1, c_2$ and $c_3$, respectively, resulting in the following effort model:
\begin{equation}
    f_{\text{effort}}(r_{\text{energy}}, r_{\text{jerk}}, r_{\text{work}}) = \frac{c_1 r_{\text{energy}} + c_2 r_{\text{jerk}} + c_3 r_{\text{work}}}{c_1 + c_2 + c_3} \tag{EJK} \label{EJK}
\end{equation}

Furthermore, we consider the three effort models from~\cite{klar_simulating_2023} (\textit{DC}, \textit{CTC}, and \textit{JAC}), where their suitability to predict mid-air pointing movements using a non-RL optimization method (MPC) was examined.
All three models include a penalty for large muscle stimulation commands %
$u$, motivated by the fact that humans seek to minimize their control effort during movement~\cite{todorov_optimal_2002}.
In the following models, this muscle effort term is penalized in the norm, whereas the \ref{EJK} model considers its mean in $r_\text{energy}$.
The \textit{DC} effort model only consists of this penalty term, weighted by a coefficient $c_1$:
\begin{equation}
    f_{\text{effort}}(u) = c_1 \|u\|^2 \tag{DC} \label{DC}
\end{equation}

The \textit{CTC} model adds a penalty on large changes in commanded torque $\tau$, which is the torque at the joints that directly results from the controlled muscle activations.
This term is motivated by a study from \citeauthor{wada_quantitative_2001}, where the minimum commanded torque \textit{change}, i.e., the derivative of $\tau$, criterion was found to be the most effective in explaining the temporal characteristics of actual hand trajectories. %
The \textit{CTC} model is described as follows:
\begin{equation}
    f_{\text{effort}}(u, \dot{\tau}) = c_1\|u\|^2 + c_2 \|\dot{\tau}\|^2 \tag{CTC} \label{CTC}
\end{equation}

Similarly, the \textit{JAC} model adds a penalty on large joint accelerations  $x_\text{qacc}$, thus avoiding "jerky" movements. This effort term was introduced in \cite{wada_quantitative_2001} and later found to provide the most comprehensive explanation of mid-air pointing movements~\cite{klar_simulating_2023}. 
In contrast to the \ref{EJK} effort model, this model penalizes the acceleration values themselves instead of their changes. 
The resulting \textit{JAC} model is defined as follows:
\begin{equation}
    f_{\text{effort}}(u, x_{\text{qacc}}) = c_1\|u\|^2 + c_2\|x_{\text{qacc}}\|^2 \tag{JAC} \label{JAC}
\end{equation}

\section{Results}
This section presents the results of training models with various reward functions. 
A qualitative evaluation of model behaviours, based on evaluation videos, complements the quantitative success rates and completion times shown in Figure~\ref{fig:success_rate}. 
The parameter values for all considered conditions (denoted as IDs in the following) are detailed in Appendix~\ref{sec:appendix}.
For a visual representation of these results, we refer to the video figure attached in the supplementary material. 

\subsection{Qualitative Results}

\begin{figure*}[h!]
    \begin{minipage}[t]{0.33\textwidth}
        \centering
        \includegraphics[width=\linewidth]{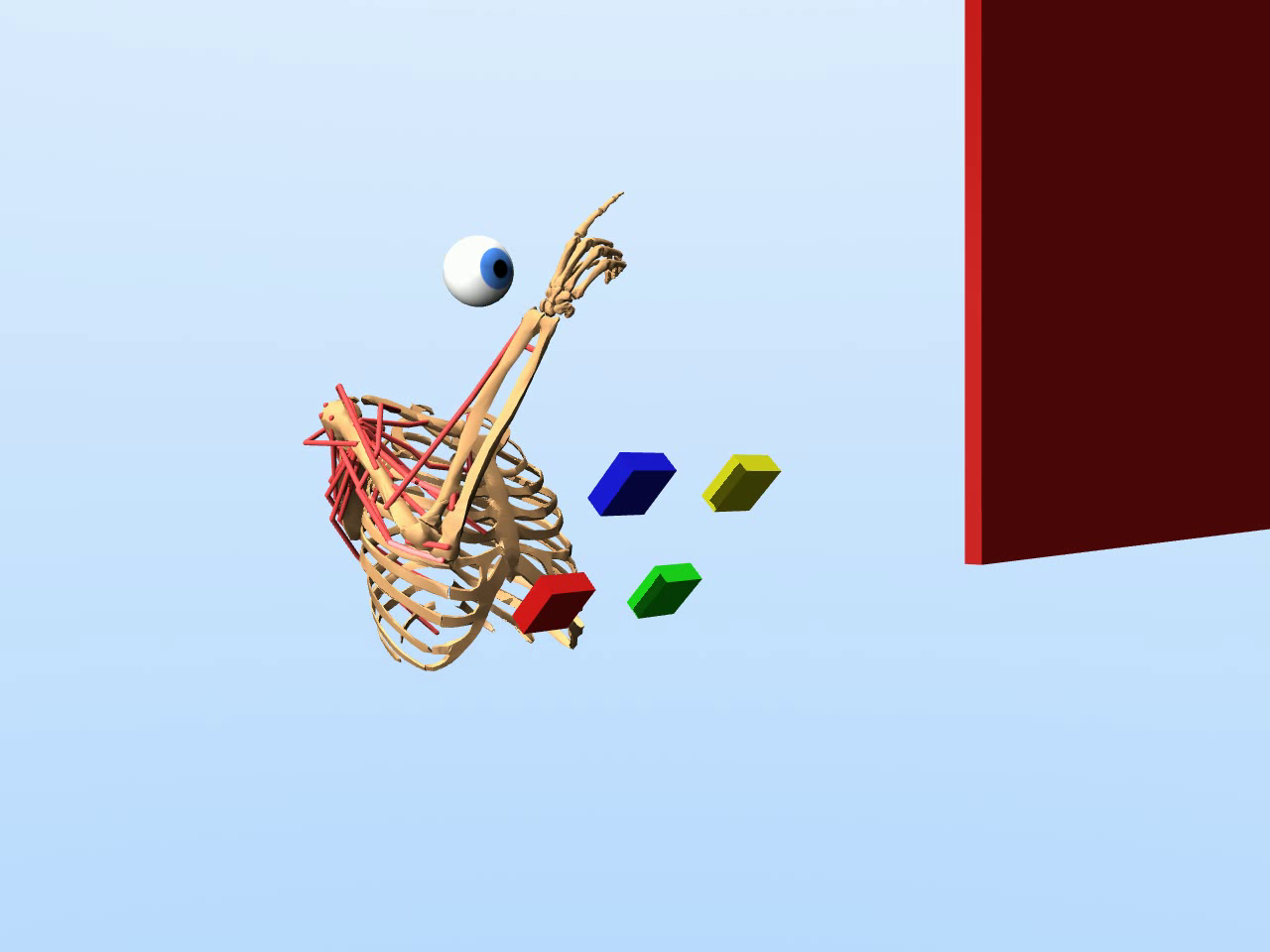}
    \end{minipage}%
    \hfill 
    \begin{minipage}[t]{0.3\textwidth}
        \centering
        \includegraphics[width=0.9\linewidth]{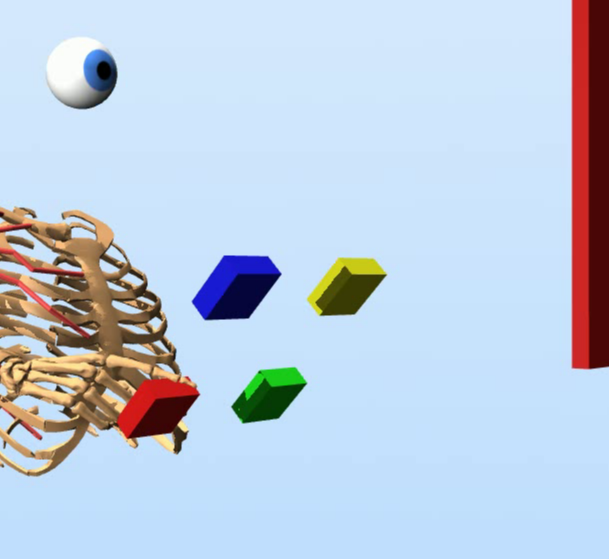}
    \end{minipage}%
    \hfill
    \begin{minipage}[t]{0.3\textwidth}
        \centering
        \includegraphics[width=0.9\linewidth]{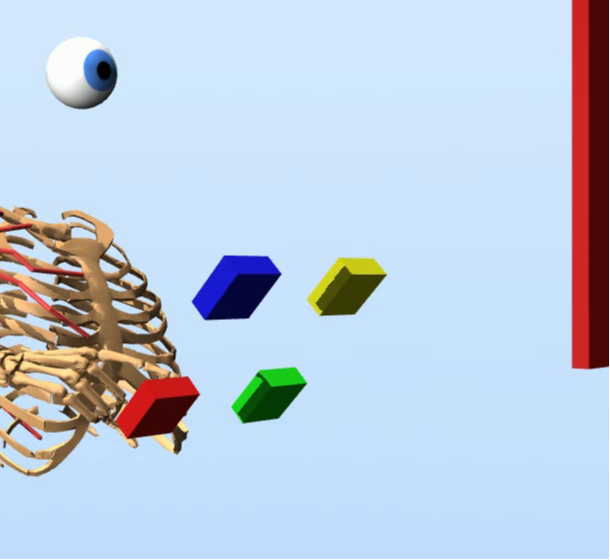}
    \end{minipage}%
    \caption{Comparison of movement patterns: only including the completion bonus leads to arbitrary movements (left); a combination of distance and effort rewards may incentivize hitting the button from the side or stopping immediately below the target (middle); a combination of completion bonus and distance rewards leads to reasonable arm movements and successful button clicks (right).
    }
    \label{fig:sparse_rewards}
    \Description{A sequence of three images illustrates different movements of a human biomechanical model
    that is positioned in front of four different coloured buttons and a display that represents one of these colours. The model represents an upper body, a right arm and an eye. In the leftmost image, the arm of the model is directed above and not towards the buttons. In the middle image, the arm reaches towards the buttons but touches the button on the upper side instead of the front. In the rightmost image, the model presses one button correctly.}
\end{figure*}

Models trained with task completion bonus only
did not learn to press all four buttons equally (ID: \ref{id31}). Instead, only the green button is successfully reached. For the remaining three colours, noisy and non-directed arm movements are generated, as shown in Figure~\ref{fig:sparse_rewards} (left). 
Increasing the bonus value %
did not lead to fundamentally different movement trajectories (ID: \ref{id32}). %

As a next step, we added a distance reward term. As expected, the distance reward helps "guiding" the RL policy towards states in which the fingertip is close to the desired button, i.e., the agent learns to identify the correct button and moves towards it.
The choice of the distance reward function has an impact on the learned strategy.
For the squared and exponential distance function, the simulated user tries to press the red button with the proximal phalanx of the index finger, i.e., the lower part of the finger close to the back of the hand, which often requires multiple attempts (IDs: \ref{id25},\ref{id27}). This behaviour was not observed for the absolute distance (ID: \ref{id26}); here, the red button is regularly (and most of the time successfully) pressed with the fingertip (see Figure \ref{fig:sparse_rewards} (right)). The remaining three buttons are regularly approached with the fingertip independent of the chosen distance function.

Using distance and effort rewards only, i.e., omitting the completion bonus in the reward function, does not lead to successful movements (IDs: \ref{id4}-\ref{id6}, \ref{id10}-\ref{id12}, \ref{id16}-\ref{id18}, \ref{id22}-\ref{id24}).
Specifically, the model hits the buttons sideways instead of pressing them correctly, consequently failing to fulfil the task, see Figure \ref{fig:sparse_rewards} (middle).

Finally, we were interested in the effects of including an effort model. %
For the \ref{CTC} effort model, we again observe a difference between the absolute distance model and squared or exponential distance; for the former, the model successfully presses all buttons, while for the latter two, the trained model is unable to press the upper two of the buttons and remains at the lower two buttons instead (IDs: \ref{id13}-\ref{id15}).
For the \ref{DC} effort model with squared or exponential distance rewards, similar behaviour as without effort term was observed, i.e., the red button specifically is pushed with the back of the hand (IDs: \ref{id19},\ref{id21}).
For the remaining effort models, i.e. \ref{JAC} and \ref{EJK}, all three distance terms result in reasonable behaviour for each button (IDs: \ref{id1}-\ref{id3}, \ref{id7}-\ref{id9}).
When the completion bonus is omitted (i.e., distance and effort rewards only (IDs: \ref{id4}-\ref{id6}, \ref{id10}-\ref{id12}, \ref{id16}-\ref{id18}, \ref{id22}-\ref{id24})), the effort models \ref{JAC}, \ref{CTC} and \ref{DC} demonstrated superior performance with the absolute value of the distance added directly, although they are only able to hit the buttons on the side. With the exponential or squared form of the distance rewards, the models are unable to hit the buttons, even from the side, and struggle with hitting the yellow button, which is the furthest away.

\begin{figure*}[h!]
    \centering
    \begin{minipage}[t]{0.246\textwidth}
        \centering
        \includegraphics[width=\textwidth]{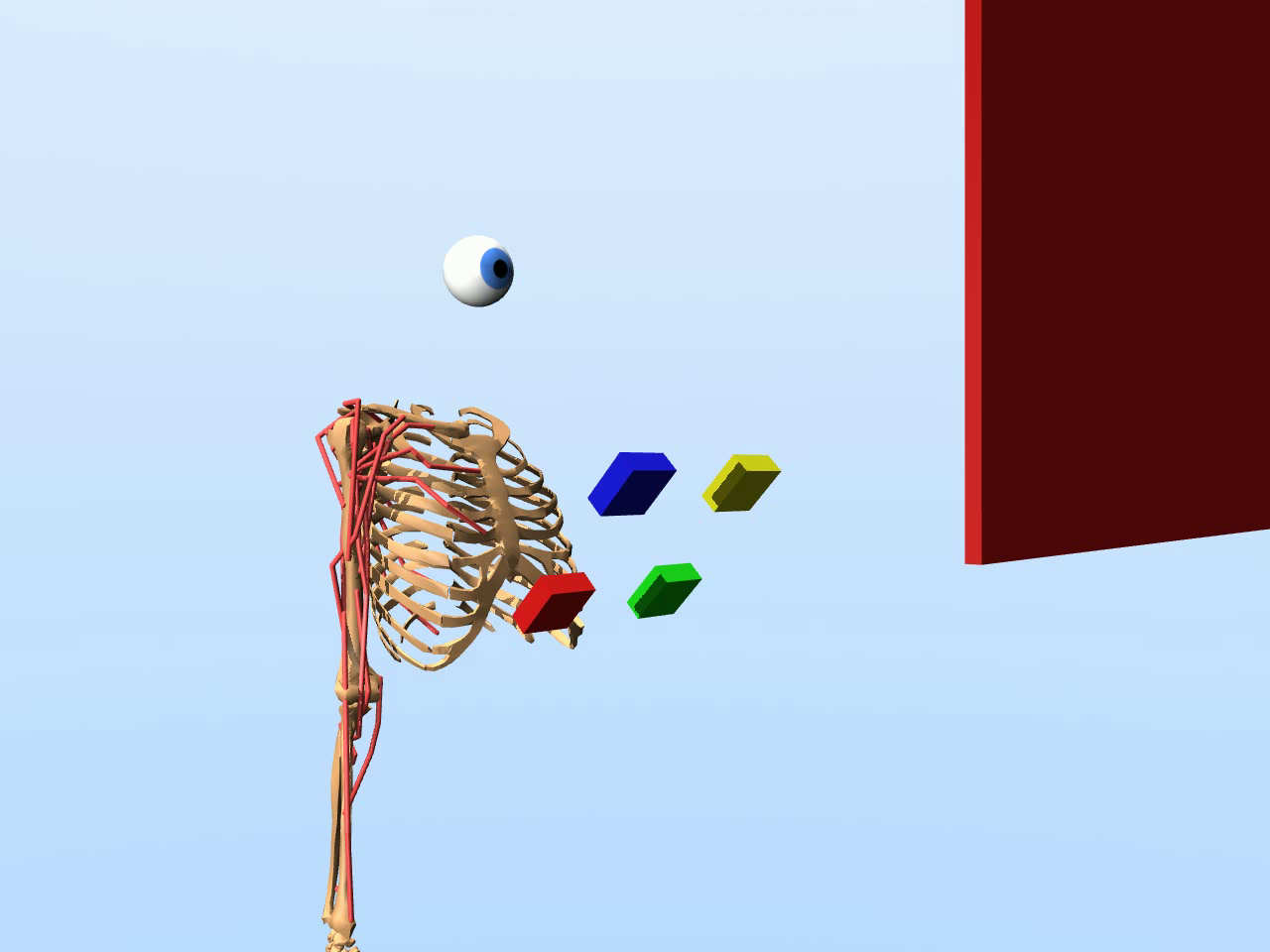}
    \end{minipage}%
    \hspace{0.01em}
    \begin{minipage}[t]{0.246\textwidth}
        \centering
        \includegraphics[width=\textwidth]{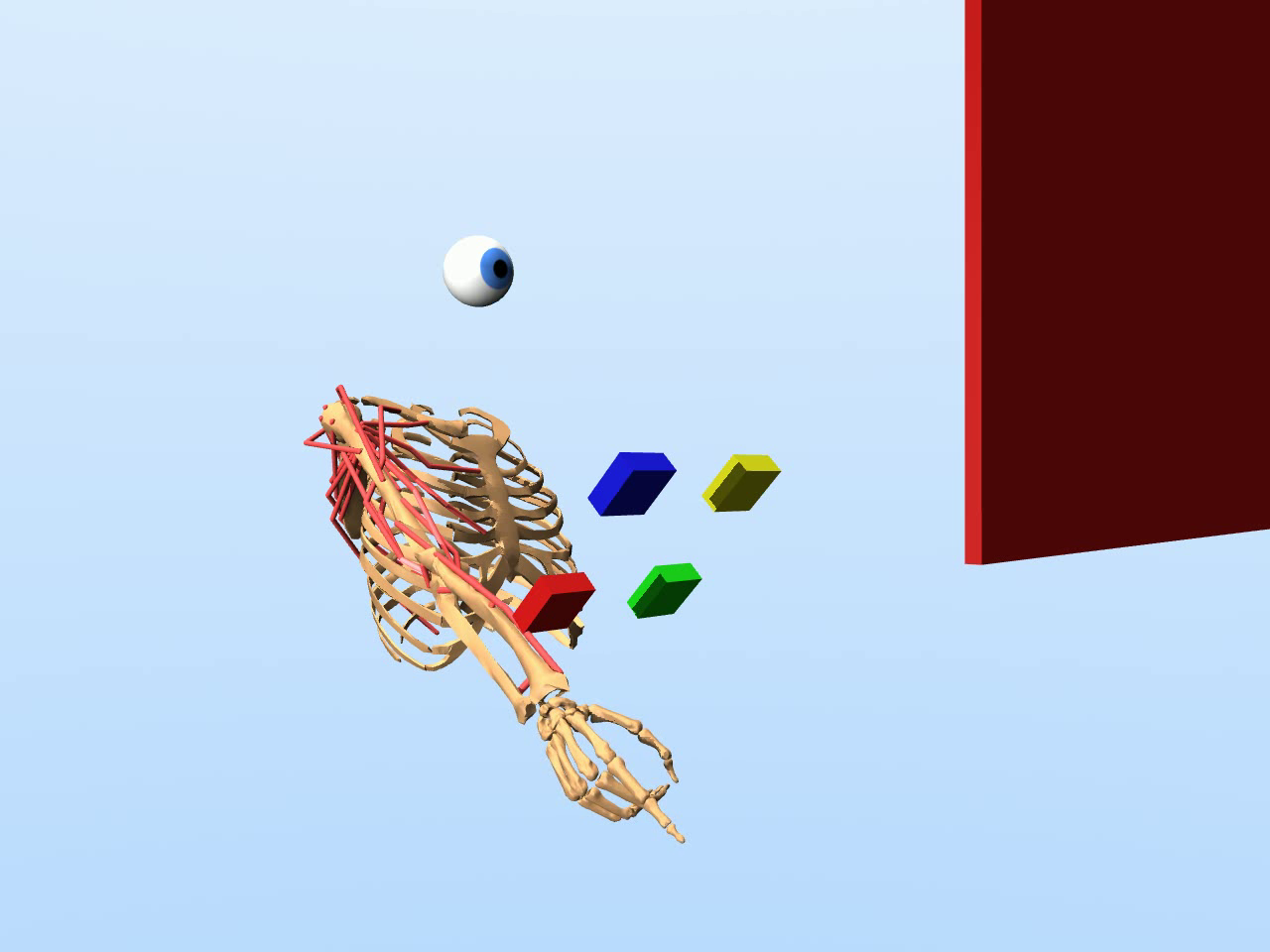}
    \end{minipage}%
    \hspace{0.01em}
    \begin{minipage}[t]{0.246\textwidth}
        \centering
        \includegraphics[width=\textwidth]{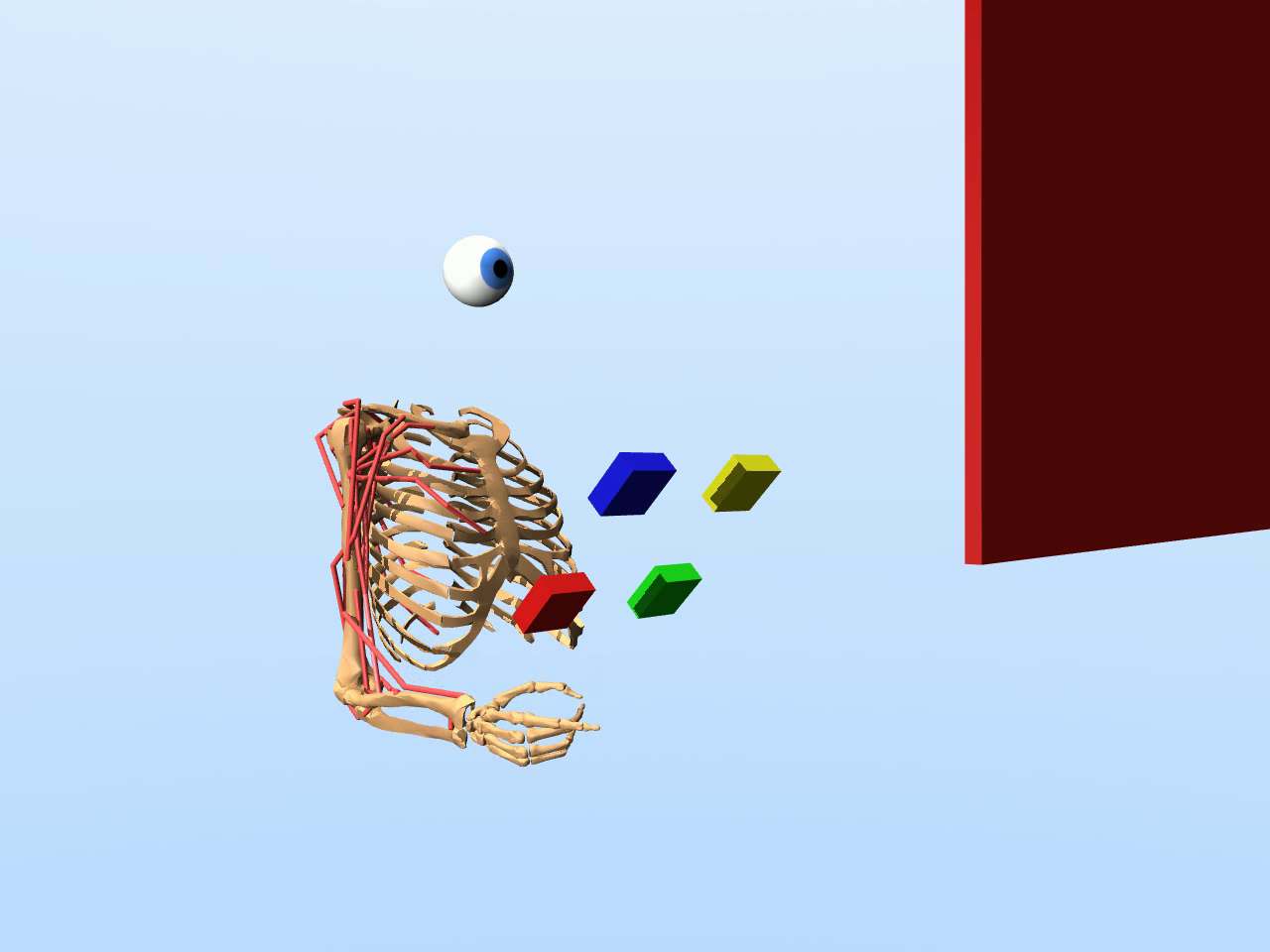}
    \end{minipage}%
    \hspace{0.01em}
    \begin{minipage}[t]{0.246\textwidth}
        \centering
        \includegraphics[width=\textwidth]{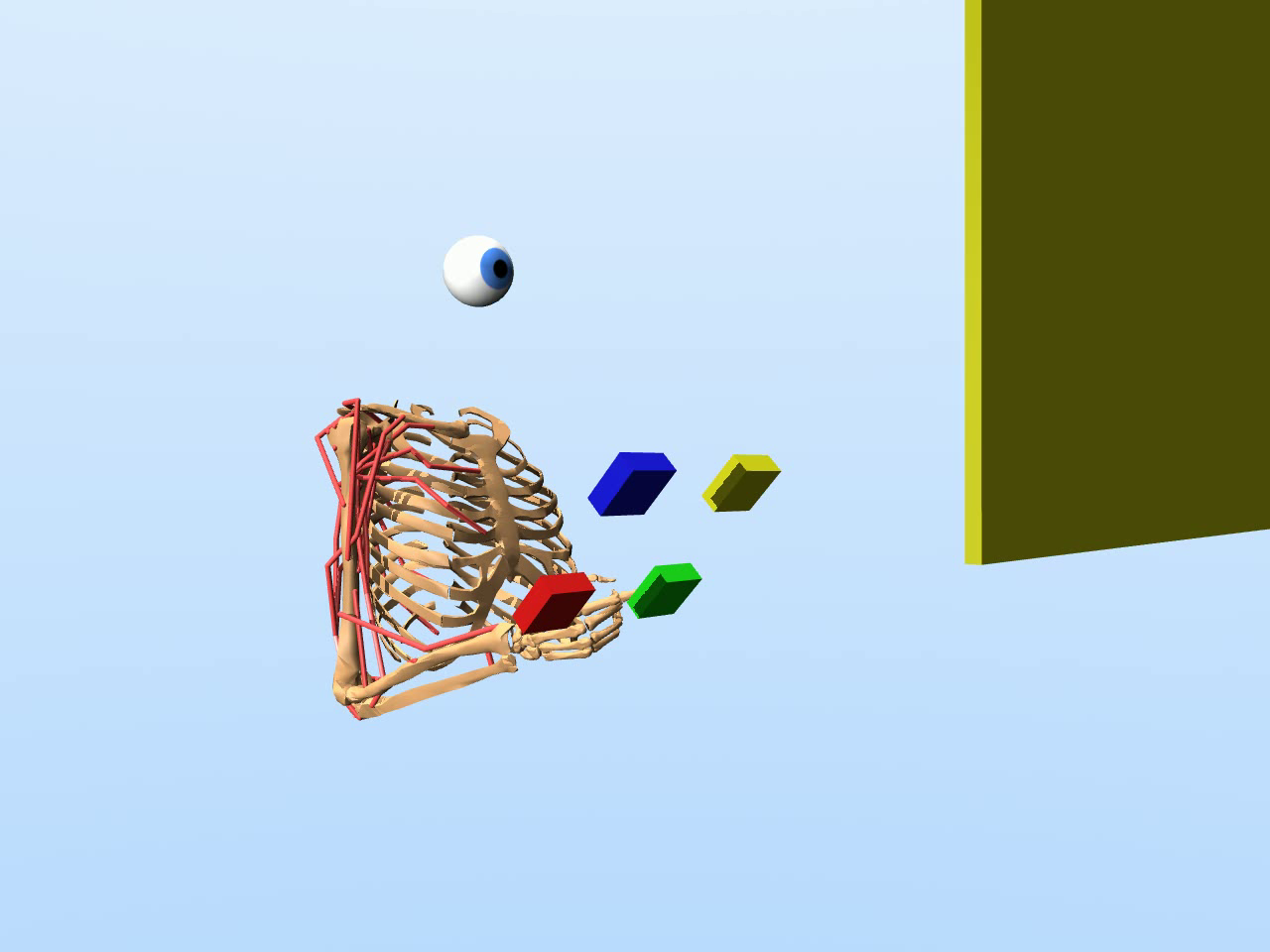}
    \end{minipage}
    \caption{Comparison of movement patterns of different effort models, from left to right: CTC model with no movement, JAC model with extended arm, DC model with bent arm, and EJK model remaining on the lower buttons.
    }
    \label{fig:effort_models}
    \Description{A sequence of four images illustrates different movement patterns of a human biomechanical model that is positioned in front of four buttons of different colours, with a display representing one of these colours.
    The model represents an upper body, a right arm and one eye.
    In the leftmost image, the arm is just hanging down. In the figure positioned adjacent to it, the arm is outstretched and pointing to a point behind the buttons.
    The figure next to this displays a human model that bends its arm in an angle of approximately 90 degrees, with the hand situated immediately beneath the buttons.
    In the final image, the human model bends its arm once more, placing its hand on the left button.}
\end{figure*}

Without a completion bonus and using exponential distance rewards (IDs: \ref{id4},\ref{id22}), the \ref{DC} and \ref{EJK} models result in the simulated user pressing nearby buttons from the side (the same applies to the \ref{JAC} and \ref{CTC} effort models with linear distance). The \ref{JAC} model with exponential distance rewards exhibits a strategy of hitting buttons from below unless the next button is directly underneath (ID: \ref{id10}).
With increased effort weights, distinctions become more apparent, as can be inferred from Figure~\ref{fig:effort_models} (IDs: \ref{id45},\ref{id53},\ref{id49}).
The \ref{JAC} model causes the arm to remain extended, whereas the \ref{DC} model bends the arm and raises it towards the buttons. The \ref{CTC} effort generates minimal rotational movements. %
However, when a completion bonus is added, most visual differences diminish, except that the \ref{DC} model causes the hand to rotate when pressing the green button.

A reward function combining only effort models (e.g., \ref{EJK}) and the completion bonus fails to initiate movement, further emphasizing the need for a complementary, task-specific "guidance" term in the reward function (IDs: \ref{id33}-\ref{id35}).

In addition, we found that the choice of the effort weight is critical for task performance. 
For example, with the \ref{EJK} effort model, exponential distance term and completion bonus, large effort weights prevent the model from pressing all buttons, limiting it to those closest to the initial position (ID: \ref{id36}).
Reducing the effort weight enables the model to press more buttons, eventually achieving full task completion.
However, further reductions lead to inconsistent performance, with increased failed attempts on the red button (ID: \ref{id43}).
All these effects were observed independently of the selected effort model.
In addition, the \ref{JAC} effort model is also sensitive towards the relative scaling of the two effort components (i.e., the choice of $c_1$ and $c_2$ in~\ref{JAC}).
Increasing the weight for the joint acceleration costs $c_2$ results in movements where the hand is placed close to the centre of the four buttons and only one of the four buttons is hit successfully (whereas, with default weight, the simulated user is able to hit all buttons (IDs: \ref{id46},\ref{id47})).

\subsection{Quantitative Results}
\begin{figure}[htpb]
    \centering
    \begin{minipage}[t]{0.49\textwidth}
        \centering
        \includegraphics[width=0.95\linewidth]{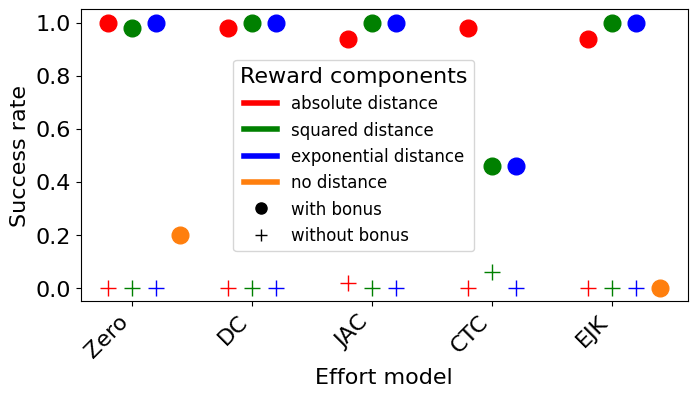}
    \end{minipage}%
    \hspace{0.01em}
    \begin{minipage}[t]{0.49\textwidth}
        \centering
        \includegraphics[width=\linewidth]{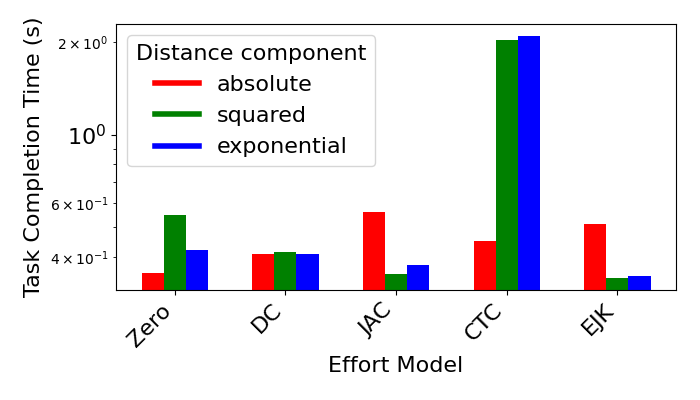}
    \end{minipage}%
    \caption{%
    Success rates (top) and average task completion times (bottom) of models trained with different reward functions of type~\eqref{eq:reward-fct-composite}.
    Full parameter details are given in Table~\ref{tab:parameters_figure} in the appendix.
    Orange circles correspond to reward functions without distance rewards and with different bonus values (1, 8, and 50, all leading to the same success rate for a given effort model). %
    The bottom figure shows the average task completion times of all models with completion bonus and a success rate of at least 50\%.
    If a model does not manage to press a button within the time limit, the maximum time of four seconds is taken.
    }
    \label{fig:success_rate}
    \Description{The figure is divided into two plots. 
    The first plot compares the success rates for the presented effort models (Zero, DC, JAC, CTC, EJK), different distance terms (absolute, squared, exponential) and the addition of a completion bonus (with, without).
    The x-axis denotes the different effort models, the y-axis the success rate from zero to one, and colours represent the different distance terms.
    Furthermore, the models that were trained with a completion bonus are represented by points, while those trained without this bonus are represented by pluses. The models trained without a completion bonus exhibit a success rate of zero for the majority of combinations, and a success rate below 20\% for two exceptions (JAC with absolute distance, and CTC with squared distance).
    In contrast, the combinations incorporating a completion bonus demonstrate a success rate that exceeds 95\%, with the exception of the models trained with the CTC effort model and squared or exponential distance, which exhibit a success rate of 50\%.The figure further depicts that the success rate of the Zero effort model with a completion bonus but without a distance reward is 20\%, and that the success rate of the EJK model with a completion bonus without a distance reward is 0\%.
    
    The second plot presents the task completion times for the combinations that achieved a success rate greater than 50\%. The y-axis represents task completion time in seconds on a logarithmic scale and the x-axis denotes the effort models.
    It is evident that the CTC effort model with the squared and exponential distance terms has the highest task completion time (approximately two seconds) in comparison to the EJK effort model with the same distance terms having the smallest task completion times (approximately 0.2 seconds).
    Furthermore, the task completion time for the DC effort model remains consistent across different distance terms, ranging from 0.3 to 0.5 seconds. This is in contrast to the other effort models, where the task completion time varies for different distance components.}
\end{figure}

Figure \ref{fig:success_rate} shows the success rates and average task completion times of 35 trained policies (see Table~\ref{tab:parameters_figure} in the appendix), calculated from 5 episodes with 10 required button clicks each.
It is evident that models trained without the completion bonus (pluses) consistently fail to achieve the task, regardless of the chosen effort model and distance term.
Adding the bonus term into the reward function improves performance significantly.
However, even with completion bonus, models that lack the distance component are unable to achieve a success rate higher than $25\%$ (only tested for EJK and zero effort).
Among the considered three distance models, the exponential distance term demonstrates the highest success rates for most conditions.
An exception is observed with the \ref{CTC} effort model, where the absolute distance term performs considerably better than squared and exponential distance rewards.

We also analyse the time the model needs to press the button.
Figure \ref{fig:success_rate} (right) demonstrates that the completion times vary between the choice of the effort model and the distance. 
The zero effort model is not always the fastest, even when considering 100\% success rates. 
Instead, the results depend on the chosen distance function. 

\section{Discussion}

\subsection{Discussion of Results and Guidelines}
The results from our simulation study demonstrate that the task completion bonus is essential; all models trained without this bonus consistently failed to complete the task. 
They usually adopted a strategy of touching the buttons from the side. 
While this strategy maximizes the distance reward, it does not result in early termination of the episode, thus leading to a suboptimal total episode return. %
We thus suggest: 
\begin{enumerate}%
    \item[\textbf{G1}] 
    Include a task-specific completion bonus as the fundamental element of the reward function.
\end{enumerate}

We also demonstrate that 
for the choice reaction task, it is not sufficient to choose a reward function consisting of solely sparse rewards. 
Adding an effort term did not improve performance, whereas including a distance reward term to guide the agent led to task completion when combined with a bonus term. 
While without an effort model, the absolute distance works best, in general, the precise choice of the distance function for best performance depends on the effort model. 
For the \ref{CTC} effort model combined with a bonus term, the absolute distance achieves superior performance.
We noticed that the absolute distance reward exhibits higher values at the beginning of the movement when compared to the other distance terms.
Since with other distances the agent does not move towards the new button but instead rests on the previous one, we suspect that the absolute distance provided "just enough" incentive.
We suggest:
\begin{enumerate}%
   \item[\textbf{G2}] %
   For complex scenarios involving vision or challenging tasks, integrate additional guidance components, such as distance-based terms, to guide the agent towards task completion.
\end{enumerate}

In addition, our results suggest that an effort term is not necessary to generate %
successful movement trajectories. %
It is important to note that humans are capable of performing the task with different body poses.
Consequently, it is possible that while the outcome remains constant, there is variation in movement pattern~\cite{guigon_computational_2007}.
This problem can be solved by including different effort terms~\cite{guigon_computational_2007}.
However, we did not identify any unreasonable patterns in the absence of the effort term. 
Our observations confirm earlier findings from \cite{fischer2021reinforcement}, where a torque-actuated model was successfully trained to point in mid-air using a completion bonus only. 
The predicted movements in~\cite{fischer2021reinforcement} even exhibit well-established movement characteristics such as Fitts' Law and the 2/3 Power Law.
However, this was mainly attributed to an adaptive target size curriculum that effectively used state exploration to ``guide'' the RL agent towards reasonable target regions. 
It will be interesting to further explore to which extent this effect can be achieved through appropriate dense reward terms instead of curriculum learning.
Since our qualitative and quantitative analysis did not indicate that implausible behaviour emerges without effort terms, 
based on the findings from our simulation study (see Figure~\ref{fig:success_rate}), we suggest:
\begin{enumerate}%
    \item[\textbf{G3}] Try without effort terms first and add one if instabilities occur. When adding effort terms, ensure comparability across different guidance components by normalizing their values and adjusting their weights where needed. 
\end{enumerate}
If an effort term is included, its weight can have a decisive effect on simulation quality.
Excessively large effort weights prevent the model from moving at all, as the effort cost outweighs the incentive to act.
Reducing effort weights incrementally enables progressive task completion in our experiments, from pressing a single button to hitting multiple targets. %
Conversely, assigning an effort weight that is too low diminishes its influence on the model's movements, possibly leading to an increase in unstable movements.
We thus suggest:
\begin{enumerate}%
    \item[\textbf{G4}] Adjust effort weights dynamically---decrease them if the model struggles to complete the task; increase them if movement instability is observed.
\end{enumerate}

\subsection{Limitations and Future Work}
While our findings provide novel insights into the intricacies of reward functions and their effects on RL-based simulated users, our work is subject to several limitations.
Our study focuses on a single interaction task, namely choice reaction, and a single biomechanical model. 
Consequently, future work should consider additional HCI tasks, such as pointing, tracking or keyboard typing \cite{ikkala_breathing_2022, hetzel_complex_interaction_2021}, and analyse the robustness of the considered and proposed reward functions to changes in the environment, task, and user model. 
This analysis could benefit from additional metrics, such as the total work done, and statements on robustness could be underpinned with statistical tests. %

While the choice and number of trained policies (60) yielded valuable insights regarding plausibility (\ref{item:rq:plausible-movement}), we only offer preliminary insights into sensitivities (\ref{item:rq:sensititivies}). 
We find that weights can be too large or too small, i.e., a sweetspot is needed. 
Additional training is required to provide more elaborate guidance on how to find that sweetspot, especially reward functions with multiple tunable weights. %

Moreover, a more profound understanding of potential biases introduced by ``shaping'' reward terms, e.g., towards certain strategies such as extending the arm early during the movement, is needed. On a technical level, separating between sensory perception and motor control in the neural network structure instead of learning visuomotor using a single neural network (i.e., end-to-end) could enhance further analysis of RL-based biomechanical simulations.

Finally, the relation between reward function tuning (e.g., adding guidance costs) and established techniques to enhance the RL training process, such as adaptive automated curriculums~\cite{fischer_sim2vr_2024, fischer2021reinforcement} or muscle-specific state exploration techniques~\cite{schumacher_dep-rl_2022, chiappa2023latent, berg2024sar}, are open questions. While more evidence is needed, our initial analysis suggests that distance reward components have the potential to restrict the motor control space to biomechanically plausible regions. %

\section{Conclusion}
Reward function design plays a crucial role for RL-based biomechanical simulations. Using a choice reaction task as a test-bed, we have analysed the individual and combined effects of three essential reward function components, namely task completion, target proximity, and effort terms. Our simulation study reveals that a combination of sparse completion bonus and dense proximity rewards is essential for task success. 
Interestingly, effort terms are dispensable if appropriate proximity rewards are used; otherwise, they need to be carefully weighted. 
Our work emphasises the need for a better understanding of the subtleties involved in training musculoskeletal models, for a variety of interaction tasks. By providing guidelines and first principles for reward function design, this work contributes towards the use of RL-based user simulations as a practical tool for HCI research and design.

\begin{acks}
This work was supported by EPSRC grant EP/W02456X/1.
Hannah Selder and Arthur Fleig acknowledge the financial support by the Federal Ministry of Education and Research of Germany and by Sächsische Staatsministerium für Wissenschaft, Kultur und Tourismus in the programme Center of Excellence for AI-research „Center for Scalable Data Analytics and Artificial Intelligence Dresden/Leipzig“, project identification number: ScaDS.AI.
\end{acks}

\bibliographystyle{ACM-Reference-Format}
\bibliography{sample-base}

\appendix
\section{Appendix}\label{sec:appendix}

The following tables present the reward functions used to train the policies.
We recall from~\eqref{eq:reward-fct-composite} that a reward function consists of the bonus, distance, and effort term and therefore amounts to
\begin{equation*}
    r_t = w_{\text{bonus}} \cdot f_{\text{bonus}}(\cdot) - w_{\text{distance}} \cdot f_{\text{distance}}(\cdot) - w_{\text{effort}} \cdot f_{\text{effort}}(\cdot).
\end{equation*}
The coefficients of the effort models are based on the results in \cite{klar_simulating_2023, charaja_generating_2024}.
For instance, the reward function of the run with ID \ref{id1} is:
\begin{equation*}
r_t = 
    \begin{cases}
        8 - \frac{r_{\text{energy}} + 8 \cdot r_{\text{jerk}} + r_{\text{work}}}{10} , & \text{correct button pressed, } \\
        0 - \frac{1-e^{-10\cdot dist}}{10} - \frac{r_{\text{energy}} + 8 \cdot r_{\text{jerk}} + r_{\text{work}}}{10}, & \text{else}.
    \end{cases}
\end{equation*}

\begin{table*}[hb]
\caption{Parameters for trained policies with distance weight $w_{\text{distance}}=1$ and bonus weight $w_{\text{bonus}}=1$.}
  \label{tab:parameters_figure}
\begin{tabular}{llllll}
\toprule
ID & Effort & $w_{\text{effort}}$    & Effort coefficients      & Distance & \ref{eq:bonus} $b$ \\
\midrule

\newtag{1}{id1} & \ref{EJK}          & 0.8 &$c_1$=1, $c_2$=8, $c_3$=1 & \ref{eq:exp_dist}      & 8   \\
\newtag{2}{id2} & \ref{EJK}          & 0.8 &$c_1$=1, $c_2$=8, $c_3$=1 & \ref{eq:absolute_dist}      & 8   \\
\newtag{3}{id3} & \ref{EJK}          & 0.8 &$c_1$=1, $c_2$=8, $c_3$=1 & \ref{eq:squared_dist}      & 8   \\
\newtag{4}{id4} & \ref{EJK}          & 0.8 &$c_1$=1, $c_2$=8, $c_3$=1 & \ref{eq:exp_dist}      & 0   \\
\newtag{5}{id5} & \ref{EJK}          & 0.8 &$c_1$=1, $c_2$=8, $c_3$=1 & \ref{eq:absolute_dist}      & 0   \\
\newtag{6}{id6} & \ref{EJK}          & 0.8 &$c_1$=1, $c_2$=8, $c_3$=1 & \ref{eq:squared_dist}      & 0   \\
\newtag{7}{id7} & \ref{JAC}          & 0.1& $c_1$=0.0198, $c_2$=$6.67\cdot10^{-5}$ & \ref{eq:exp_dist}      & 8   \\
\newtag{8}{id8} & \ref{JAC}          & 0.1& $c_1$=0.0198, $c_2$=$6.67\cdot10^{-5}$ & \ref{eq:absolute_dist}      & 8   \\
\newtag{9}{id9} & \ref{JAC}          & 0.1& $c_1$=0.0198, $c_2$=$6.67\cdot10^{-5}$ & \ref{eq:squared_dist}      & 8   \\
\newtag{10}{id10} & \ref{JAC}          & 0.1& $c_1$=0.0198, $c_2$=$6.67\cdot10^{-5}$ & \ref{eq:exp_dist}      & 0   \\
\newtag{11}{id11} & \ref{JAC}          & 0.1& $c_1$=0.0198, $c_2$=$6.67\cdot10^{-5}$ & \ref{eq:absolute_dist}      & 0   \\
\newtag{12}{id12} & \ref{JAC}          & 0.1& $c_1$=0.0198, $c_2$=$6.67\cdot10^{-5}$ & \ref{eq:squared_dist}      & 0   \\
\newtag{13}{id13} & \ref{CTC}          & 0.01& $c_1$=0.649, $c_2$=0.0177 & \ref{eq:exp_dist}      & 8   \\
\newtag{14}{id14} & \ref{CTC}          & 0.01& $c_1$=0.649, $c_2$=0.0177 & \ref{eq:absolute_dist}      & 8   \\
\newtag{15}{id15} & \ref{CTC}          & 0.01& $c_1$=0.649, $c_2$=0.0177 & \ref{eq:squared_dist}      & 8   \\
\newtag{16}{id16} & \ref{CTC}          & 0.01& $c_1$=0.649, $c_2$=0.0177 & \ref{eq:exp_dist}      & 0   \\
\newtag{17}{id17} & \ref{CTC}          & 0.01& $c_1$=0.649, $c_2$=0.0177 & \ref{eq:absolute_dist}      & 0   \\
\newtag{18}{id18} & \ref{CTC}          & 0.01& $c_1$=0.649, $c_2$=0.0177 & \ref{eq:squared_dist}      & 0   \\
\newtag{19}{id19} & \ref{DC}          & 0.01& $c_1$=0.1477 & \ref{eq:exp_dist}      & 8   \\
\newtag{20}{id20} & \ref{DC}          & 0.01& $c_1$=0.1477 & \ref{eq:absolute_dist}      & 8   \\
\newtag{21}{id21} & \ref{DC}          & 0.01& $c_1$=0.1477 & \ref{eq:squared_dist}      & 8   \\
\newtag{22}{id22} & \ref{DC}          & 0.01& $c_1$=0.1477 & \ref{eq:exp_dist}      & 0   \\
\newtag{23}{id23} & \ref{DC}          & 0.01& $c_1$=0.1477 & \ref{eq:absolute_dist}      & 0   \\
\newtag{24}{id24} & \ref{DC}          & 0.01& $c_1$=0.1477 & \ref{eq:squared_dist}      & 0   \\
\newtag{25}{id25} & Zero          & 0 &  & \ref{eq:exp_dist}      & 8   \\
\newtag{26}{id26} & Zero          & 0 &  & \ref{eq:absolute_dist}      & 8   \\
\newtag{27}{id27} & Zero          & 0 &  & \ref{eq:squared_dist}      & 8   \\
\newtag{28}{id28} & Zero          & 0 &  & \ref{eq:exp_dist}      & 0   \\
\newtag{29}{id29} & Zero          & 0 &  & \ref{eq:absolute_dist}      & 0   \\
\newtag{30}{id30} & Zero          & 0 &  & \ref{eq:squared_dist}      & 0   \\
\newtag{31}{id31} & Zero          & 0 &    &   & 8   \\
\newtag{32}{id32} & Zero          & 0 &    &   & 50   \\
\newtag{33}{id33} & \ref{EJK}          & 0.8 & $c_1$=1, $c_2$=8, $c_3$=1&       & 1   \\
\newtag{34}{id34} & \ref{EJK}          & 0.8 & $c_1$=1, $c_2$=8, $c_3$=1&       & 8   \\
\newtag{35}{id35} & \ref{EJK}          & 0.8 & $c_1$=1, $c_2$=8, $c_3$=1&       & 50   \\

\bottomrule
\end{tabular}
\end{table*}

\begin{table*}[]
\caption{Further trainings with exponential distance, distance weight $w_{\text{distance}}=1$, and bonus weight $w_{\text{bonus}}=1$}.
  \label{tab:parameters_exp_dist}
\begin{tabular}{llllll}
\toprule
ID & Effort & $w_{\text{effort}}$    & Effort coefficients      & Distance & \ref{eq:bonus} $b$ \\
\midrule
\newtag{36}{id36} & \ref{EJK}          & 16 &$c_1$=1, $c_2$=8, $c_3$=1 & \ref{eq:exp_dist}      & 8   \\
\newtag{37}{id37} & \ref{EJK}          & 8 &$c_1$=1, $c_2$=8, $c_3$=1 & \ref{eq:exp_dist}      & 8   \\
\newtag{38}{id38} & \ref{EJK}          & 4 &$c_1$=1, $c_2$=8, $c_3$=1 & \ref{eq:exp_dist}      & 8   \\
\newtag{39}{id39} & \ref{EJK}          & 1.6 &$c_1$=1, $c_2$=8, $c_3$=1 & \ref{eq:exp_dist}      & 8   \\
\newtag{40}{id40} & \ref{EJK}          & 0.4 &$c_1$=1, $c_2$=8, $c_3$=1 & \ref{eq:exp_dist}      & 8   \\
\newtag{41}{id41} & \ref{EJK}          & 0.16 &$c_1$=1, $c_2$=8, $c_3$=1 & \ref{eq:exp_dist}      & 8   \\
\newtag{42}{id42} & \ref{EJK}          & 0.08 &$c_1$=1, $c_2$=8, $c_3$=1 & \ref{eq:exp_dist}      & 8   \\
\newtag{43}{id43} & \ref{EJK}          & 0.04 &$c_1$=1, $c_2$=8, $c_3$=1 & \ref{eq:exp_dist}      & 8   \\
\newtag{44}{id44} & \ref{JAC}          & 1& $c_1$=0.0198, $c_2$=$6.67\cdot10^{-5}$ & \ref{eq:exp_dist}      & 8   \\
\newtag{45}{id45} & \ref{JAC}          & 1& $c_1$=0.0198, $c_2$=$6.67\cdot10^{-5}$ & \ref{eq:exp_dist}      & 0   \\
\newtag{46}{id46} & \ref{JAC}          & 0.01& $c_1$=0.0198, $c_2$=$6.67\cdot10^{-5}$ & \ref{eq:exp_dist}      & 8   \\
\newtag{47}{id47} & \ref{JAC}          & 0.01& $c_1$=0.0198, $c_2$=$6.67\cdot10^{-4}$ & \ref{eq:exp_dist}      & 8   \\
\newtag{48}{id48} & \ref{CTC}          & 1& $c_1$=0.649, $c_2$=0.0177 & \ref{eq:exp_dist}      & 8   \\
\newtag{49}{id49} & \ref{CTC}          & 1& $c_1$=0.649, $c_2$=0.0177 & \ref{eq:exp_dist}      & 0   \\
\newtag{50}{id50} & \ref{CTC}          & 0.1& $c_1$=0.649, $c_2$=0.0177 & \ref{eq:exp_dist}      & 8   \\
\newtag{51}{id51} & \ref{CTC}          & 0.001& $c_1$=0.649, $c_2$=0.0177 & \ref{eq:exp_dist}      & 8   \\
\newtag{52}{id52} & \ref{DC}          & 1& $c_1$=0.1477 & \ref{eq:exp_dist}      & 8   \\
\newtag{53}{id53} & \ref{DC}          & 1& $c_1$=0.1477 & \ref{eq:exp_dist}      & 0   \\
\newtag{54}{id54} & \ref{DC}          & 0.001& $c_1$=0.1477 & \ref{eq:exp_dist}      & 8   \\
\newtag{55}{id55} & \ref{DC}          & 1& $c_1$=0.0001 & \ref{eq:exp_dist}      & 8   \\
\newtag{56}{id56} & \ref{DC}          & 1& $c_1$=0.0001 & \ref{eq:exp_dist}      & 8   \\
\newtag{57}{id57} & \ref{DC}          & 5& $c_1$=0.0001 & \ref{eq:exp_dist}      & 8   \\
\newtag{58}{id58} & \ref{DC}          & 10& $c_1$=0.0001 & \ref{eq:exp_dist}      & 8   \\
\newtag{59}{id59} & \ref{DC}          & 50& $c_1$=0.0001 & \ref{eq:exp_dist}      & 8   \\
\newtag{60}{id60} & \ref{DC}          & 100& $c_1$=0.0001 & \ref{eq:exp_dist}      & 8   \\

\bottomrule
\end{tabular}
\end{table*}

\end{document}